\documentstyle[11pt]{article}

\renewcommand{\thefootnote}{\fnsymbol{footnote}}

\newtheorem{theorem}{Theorem}
\newtheorem{lemma}{Lemma}
\newcommand{\eqd}{\stackrel{\triangle}{=}}
\newcommand{\dfn}{\stackrel{\triangle}{=}}

\newcommand {\bc} {\mbox{\boldmath $c$}}
\newcommand {\bd} {\mbox{\boldmath $1$}}

\newcommand {\bl} {\mbox{\boldmath $l$}}

\newcommand {\bp} {\mbox{\boldmath $p$}}

\newcommand {\br} {\mbox{\boldmath $r$}}

\newcommand {\bw} {\mbox{\boldmath $w$}}
\newcommand {\bx} {\mbox{\boldmath $x$}}

\newcommand {\bE} {\mbox{\boldmath $E$}}

\newcommand{\calI}{{\cal I}}

\newcommand{\calX}{{\cal X}}

\def\parsec{\par\noindent}
\def\big{\bigskip\parsec}
\def\med{\medskip\parsec}

\begin{document}
\thispagestyle{empty}
\setcounter{page}{1}
\setlength{\baselineskip}{1.5\baselineskip}
\title{Average Redundancy of the Shannon Code for 
Markov Sources}
\author{
Neri Merhav\thanks{N.~Merhav is with the
Department of Electrical Engineering,
Technion -- Israel Institute of Technology, Technion City, Haifa 32000,
Israel. E-mail: {\tt merhav@ee.technion.ac.il}. N.~Merhav's work was supported
by the Israeli Science Foundation (ISF), grant no.\ 208/08.}
\and Wojciech Szpankowski\thanks{W.~Szpankowski is with the
Department of Computer Science,
Purdue University,
W.~Lafayette, IN 47907, U.S.A.
E-mail: {\tt spa@cs.purdue.edu}.
He is also a Visiting Professor at ETI,  Gda\'{n}sk University of
Technology, Poland.
W.~Szpankowski's work was
supported in part by the NSF Science and Technology Center for
Science of Information Grant CCF-0939370, NSF Grant
CCF-0830140, AFOSR Grant FA8655-11-1-3076, and NSA Grant H98230-11-1-0141.}}
\maketitle


\begin{abstract}
It is known that for memoryless sources,
the average and maximal redundancy of fixed--to--variable length codes, such as
the Shannon and Huffman codes, exhibit two modes of behavior for long blocks.
It either converges to a limit or it
has an oscillatory pattern, depending on the irrationality or rationality,
respectively, of certain parameters that depend on the source. 
In this paper, we extend these
findings, concerning the Shannon code, to the case of a Markov source,
which is considerably more involved.
While this dichotomy,
of convergent vs.\ oscillatory behavior, is well 
known in other contexts (including renewal theory, ergodic 
theory, local limit theorems and large deviations of 
discrete distributions), 
in information theory (e.g., in redundancy analysis) it was
recognized relatively recently.
To the best of our knowledge, no results of this type
were reported thus far for Markov sources. We provide a precise
characterization of the
convergent vs.\ oscillatory behavior of the Shannon code redundancy for a class
of irreducible, periodic and aperiodic, Markov sources. These findings are obtained by 
analytic methods, such as Fourier/Fej\'er series analysis and spectral analysis of matrices.
\end{abstract}

\big
{\bf Index Terms:} Shannon code, average redundancy, Fourier series, uniform
convergence, spectral analysis, analytic information theory.

\newpage

\renewcommand{\thefootnote}{\arabic{footnote}}
\setcounter{footnote}{0}

\section{Introduction}

Recent years have witnessed a resurgence of interest in
redundancy rates of lossless coding, see, e.g.,
\cite{abrahams}, \cite{ds04}, \cite{Gallager78}, \cite{js04}, 
\cite{LS97}, \cite{merhav12}, 
\cite{rissanen}, \cite{Savari98}, \cite{SG97}, \cite{spa00}.
In particular,
in \cite{spa00} Szpankowski derived
asymptotic expressions of the (unnormalized) average redundancy $R_n$, 
as a function of the block length $n$, for the Shannon code, the Huffman
code, and other codes, focusing primarily on the binary memoryless source
(BSS), parametrized by $p$ -- the probability of `1'. 
A rather interesting behavior of $R_n$ was revealed in \cite{spa00}, 
especially in the cases of the
Shannon code and the Huffman code: When $\alpha\dfn \log_2[(1-p)/p]$
is irrational, then $R_n$ converges 
to a constant (which is $1/2$ for the Shannon code),
as $n\to \infty$. On the other hand,
when $\alpha$ is rational, $R_n$ has a non--vanishing oscillatory term 
whose fundamental frequency and amplitude depend on the source
statistics in an explicit manner.

More precisely, confining the discussion to the Shannon code, in
\cite{spa00} the average unnormalized redundancy
\begin{equation}
R_n = \bE\left\{\lceil -\log_2 P(X_1,\ldots,X_n)\rceil+\log_2
P(X_1,\ldots,X_n)\right\},
\end{equation}
was analyzed for large $n$, assuming that the source $P$,
that governs the data to be compressed,
$X_1,X_2,\ldots$, is a BSS. A straightforward 
extension (see also \cite{merhav12}) of
the Shannon--code redundancy result of \cite{spa00}, 
to a general $r$--ary alphabet memoryless source,
with letter probabilities $p_1,\ldots,p_r$, yields the following expression:
\begin{equation}
\label{rn}
R_n =\left\{\begin{array}{ll}
\frac{1}{2}+\frac{1}{M}\left(\frac{1}{2}-\left<\beta Mn\right>\right)
+o(1) & \mbox{all $\{\alpha_j\}$ are rational}\\
\frac{1}{2}+o(1) & \mbox{otherwise}
\end{array}\right.
\end{equation}
where $\beta\dfn-\log p_1$, $\alpha_j=\log p_j/p_1$, $j=2,3,\ldots,r$,
$\left<u\right>$ is the fractional part of a real number $u$
(i.e., $\left<u\right>=u-\lfloor u \rfloor$),
and $M$ is the smallest common multiple of all
denominators of the rational numbers $\{\alpha_j\}$ 
when presented as ratios between two relatively
prime integers. This erratic behavior, where $R_n$ is either 
convergent (and then the limit is always $1/2$) or
oscillatory, depending on the rationality of $\{\alpha_j\}$,
was related in \cite{merhav12} to wave
diffraction patterns of scattering from partially disordered media, where the
existence/non--existence of Bragg peaks depends on 
the rationality/irrationality  of certain optical distance
ratios.

Our goal in this paper 
is to extend the scope of this analysis to irreducible Markov sources and
to evaluate
precisely (for large $n$) the average redundancy of the Shannon code for
a finite alphabet, first order Markov source 
with given transition probabilities.
In doing so, we also provide a more complete analysis than in \cite{merhav12} and
\cite{spa00}.
As will be seen, 
this extension to the Markov case appears
rather non--trivial, both from the viewpoint of the conditions for oscillatory
behavior and from the aspect of the asymptotic expression of $R_n$ in the oscillatory mode.
These depend strongly on the dominant eigenvalues and on the detailed structure of
the matrix of transition probabilities. For example, in contrast to the 
memoryless case, where there is only one oscillatory term, 
when it comes to the Markov case, in the oscillatory mode
there are, in general, contributions from multiple oscillatory terms, and
in the convergent mode, $R_n$ may converge to a constant other
than $1/2$ (see Example 2 below). Moreover, it turns out that the
behavior of the redundancy depends quite
strongly on important dynamical properties of the Markov chain, such as 
reducibility/irreducibility and periodicity/aperiodicity.

We begin our study (Sections 2 and 3) 
from the relatively simple case where all single--step state transitions
have positive probability.
Our main result in Section 2, Theorem 1, is then an extension of formula (\ref{rn}) to the
Markov case with strictly positive state transition probabilities. To give the
reader a general idea of this theorem, an informal description of it can be stated
as follows: Rather than the parameters $\{\alpha_j\}$ of the memoryless case, we
now define a matrix $\{\alpha_{jk}\}_{j,k=1}^r$ of log--ratios of
certain transition probabilities (the exact definition will be provided in the
sequel). If at least one of these parameters is irrational, then similarly as
in the memoryless case, $R_n=\frac{1}{2}+o(1)$. If, on the other hand, all
these parameters are rational, then as in the memoryless case, let $M$ be their
smallest common denominator. In this case, $R_n=\Omega_n+o(1)$, for ``most
large values'' of $n$ (a term that will be defined precisely in the sequel),
where $\Omega_n$ is a linear combination of certain functions of $n$, for
which we have an explicit formula in terms of the source parameters. These
functions oscillate as $n$ varies, with amplitude $1/M$ and a fundamental frequency
that depends on the source parameters. 

In Section
4, we relax the strict positivity assumption, but still assume the Markov
chain to be irreducible. Under this assumption, we first assume that the
chain is also aperiodic, and then further extend the scope to allow periodicity.
In these cases, the extension of eq.\ (\ref{rn}) is still available, though
it is somewhat less explicit (than in the positive 
transition matrix case) in the sense that
it depends on certain parameters of the source, for which we have no closed--form
expressions, but which can be found by numerical procedures.
It is also demonstrated (in Example 2) 
that the irreducibility assumption is essential, 
since the above described two--mode behavior ceases to exist when this 
assumption is dropped.

We should point out that minimax redundancy and regret for
the {\it class} of Markov sources were studied in the past -- 
see, e.g., \cite{js04}, \cite{rissanen}. Interestingly enough, the minimax regret
for memoryless and Markov sources does not exhibit the two--mode behavior
of either convergent or oscillatory mode \cite{ds04}. 
This dichotomy, of convergent vs.\ oscillatory behavior, 
with dependence on rationality/irrationality of certain parameters, is a well
recognized phenomenon in mathematics and physics,
ranging across a large variety of
areas, including renewal
theory, ergodic theory \cite{gray}, local limit theorems and
large deviations for discrete distributions \cite{blackwell}, \cite{esseen}.
This phenomenon, however, was observed in information theory only
relatively recently \cite{gray}, \cite{spa00}. On the other hand, the oscillatory phenomenon
for discrete random structures is a well known fact in analysis
of algorithms \cite{fs-book}, \cite{spa-book}, and also in information theory
\cite{ds04}, \cite{LS97}, \cite{spa-book}.

\section{Formulation and Results for Positive Transition Matrices}
\label{sec-main}

In this section, we first 
establish notation conventions and spell out our assumptions.
Then, we present our main result for the case of a positive
transition probability matrix (Theorem 1), discuss it, and provide an example for its use.

Throughout this paper, we adopt the customary notation conventions in the
information theory literature: Random variables will be denoted by capital
letters (e.g., $X$), specific values they may take will be denoted by
the corresponding lower--case letters (e.g., $x$), and their alphabets will be
denoted by the corresponding calligraphic letters (e.g., $\calX$). Random
vectors of length $n$ 
(e.g., $(X_1,X_2,\ldots,X_n)$) 
will be denoted by
capital letters superscripted by $n$ (e.g., $X^n$), and specific values of these
vectors (e.g., $(x_1,x_2,\ldots,x_n)$) will be denoted by lower--case letters
superscripted by $n$ (e.g., $x^n$). Finally, the set of vectors of length $n$,
with components taking on values in $\calX$, will be denoted by $\calX^n$.
Logarithms will always be understood to be taken w.r.t.\ the base 2.
The function $\calI(\cdot)$ will denote the indicator function, that is,
for a given statement $E$, $\calI(E)=1$ if $E$ is true, and $\calI(E)=0$ if
$E$ is false.

Consider a source
sequence $X_1,X_2,\ldots$, $X_t\in\calX=\{1,2,\ldots,r\}$
($r$ -- positive integer), $t=1,2,\ldots$, governed by a
first--order Markov chain with a given matrix $P$
of state--transition probabilities $\{p(j|k)\}_{j,k=1}^r$.
The initial state probabilities will be denoted
by $p_k$, $k=1,2,\ldots,r$.
The stationary state probabilities will be denoted by $\pi_k$,
$k=1,2,\ldots,r$. Thus, the probability of a given source string
$x^n=(x_1,\ldots,x_n)\in\calX^n$, under the given Markov source, is
\begin{equation}
\mu(x^n)=p_{x_1}\prod_{t=2}^np(x_t|x_{t-1}).
\end{equation}
The average unnormalized redundancy of the Shannon code is defined as
\begin{equation}
\label{spa1}
R_n \dfn
\bE\{\lceil-\log \mu(X^n)\rceil +\log \mu(X^n)\},
\end{equation}
where here and throughout the sequel, $\bE\{\cdot\}$ denotes the expectation
operator w.r.t.\ the underlying Markov source $\mu$ just defined.

As mentioned in the Introduction, in this paper, we assume that
$P$ is irreducible.
We remind the reader that
an {\it irreducible} Markov chain is one where there is positive probability
to pass from every
state $j\in\calX$ to every state $k\in\calX$ within a finite number of steps,
namely, for every $j$ and $k$, there exists a positive integer $l$ such that
the $(k,j)$--th element of $P^l$ is strictly positive. 
Another important concept we will need is periodicity.
The {\it period} $d_j$
of a state $j$ is the greatest common divisor of all integers $n$ for which
$\mbox{Pr}\{X_n=j|X_0=j\}> 0$. A state is called periodic if $d_j >1$ and
aperiodic if $d_j=1$. Since all states of an irreducible Markov chain are
in the same class of communicating states, then $d_j$ is the same for all
states, and hence will be denoted collectively by $d$. An irreducible Markov
chain is then called periodic if $d > 1$ and aperiodic if $d=1$. The
case where all entries of $P$ are positive, henceforth referred to as the
case of a {\it positive matrix} $P$, is obviously a case of
an irreducible, aperiodic Markov chain. However, the positivity of $P$
is not a necessary condition for irreducibility and aperiodicity of a
Markov chain. Throughout the remaining part of this section, as well as in Section 3,
we assume that all entries of $P$ are strictly positive.

Our main result in this section is the following (the proof appears in
Section 3).
\begin{theorem}
\label{th-main}
Consider the Shannon code of block 
length $n$ for a Markov source $\mu$ with a
a given vector $\bp=(p_1,\ldots,p_r)$
of initial state probabilities and a positive 
state transition matrix $P$. Define
\begin{equation}
\alpha_{jk}=\log\left[\frac{p(j|1)p(j|j)}{p(k|1)p(j|k)}\right],~~~~j,k\in\{1,2,\ldots,r\}.
\end{equation}
Then, the redundancy $R_n$ is characterized as follows:\\
(a) If not all $\{\alpha_{jk}\}$ are rational, then
\begin{equation}
R_n =\frac{1}{2}+o(1).
\end{equation}
(b) If all $\{\alpha_{jk}\}$ are rational, then
for every $j,k\in\{1,\ldots,r\}$, let
\begin{equation}
\label{zeta}
\zeta_{jk}(n)=M[-(n-1)\log p(1|1)+\log p(j|1)-\log p(k|1)-\log p_j],
\end{equation}
and
\begin{equation}
\Omega_n=\frac{1}{2}\left(1-\frac{1}{M}\right)+
\frac{1}{M}\sum_{j=1}^r\sum_{k=1}^rp_j\pi_k\varrho[\zeta_{jk}(n)],
\end{equation}
where $\varrho(u)\dfn\lceil u\rceil -u$ and
$M$ is the smallest common integer multiple of the denominators of 
$\{\alpha_{jk}\}$, when each one of these numbers is represented
as a ratio between two relatively prime integers. Then,
there exists a positive sequence $\xi_n\to 0$, which depends only the source
parameters, such that $R_n$ is upper bounded and lower bounded as follows:
\begin{equation}
\label{e3u}
R_n\le\Omega_n+
\frac{1}{M}\sum_{j=1}^r\sum_{k=1}^rp_j\pi_k\calI\{
\varrho[\zeta_{jk}(n)]\notin(\xi_n,1-\xi_n)\}+o(1).
\end{equation}
\begin{equation}
\label{e3l}
R_n\ge\Omega_n-
\frac{1}{M}\sum_{j=1}^r\sum_{k=1}^rp_j\pi_k
\calI\{\varrho[\zeta_{jk}(n)]\notin(\xi_n,1-\xi_n)\}-o(1).
\end{equation}
\end{theorem}

As a technical comment,
it should be pointed out that the choice
of the index $1$ in the conditioning of $p(j|1)$ and
$p(k|1)$, that appear in the definition
of $\alpha_{jk}$ and in (\ref{zeta}), is completely
arbitrary. One may choose any
other index in $\{1,2,\ldots,r\}$, as long as it is the same
index in both places in the expression of $\alpha_{jk}$,
as well as in the second and third
terms in the square brackets of (\ref{zeta}).
Also, $p(1|1)$ in (\ref{zeta}) can be
replaced independently by $p(l|l)$ for any $l\in\{1,2\ldots,r\}$.\\

\noindent
{\bf Discussion.} Theorem 1 tells us that, similarly as in the memoryless case, in the positive
matrix case, $R_n$ has two
modes of behavior. In the convergent mode, which happens when at least one
$\alpha_{jk}$ is irrational, $R_n \to 1/2$. In the oscillatory mode, which
happens when all $\{\alpha_{jk}\}$  are rational, $R_n$ oscillates and it
asymptotically coincides with $\Omega_n$
for {\it most large values}\footnote{The statement ``$R_n$ asymptotically coincides with
$\Omega_n$ for most large values of $n$'' means that
for every $\epsilon> 0$,
the fraction of values of $n$, within the range $\{1,\ldots,N\}$, for which
$|R_n-\Omega_n|>\epsilon$,
tends to zero as $N\to\infty$.}
of $n$, provided that $\log p(1|1)$ is irrational. 
This follows from the following consideration: If $\log p(1|1)$ is irrational, 
then by Weyl's equidistribution theorem \cite{KN74}, the sequences
$\{\zeta_{jk}(n)\}_{n\ge 1}$ 
are uniformly distributed modulo 1, i.e., they fill
the unit interval $\mbox{mod}~1$ with a uniform density as $n$ exhausts the
positive integers. Thus, for every fixed $\xi$,
$\varrho[\zeta_{jk}(n)]\notin(\xi,1-\xi)$ for a
fraction $2\xi$ of the values of $n$. 
This means that for $\xi_n\to 0$, the terms
$\calI\{\varrho[\zeta_{jk}(n)]\notin(\xi_n,1-\xi_n)\}$ vanish for most 
large values of $n$,
and then the lower bound and the 
upper bound on $R_n$ asymptotically coincide with
$\Omega_n$. If, on the
other hand, $\log p(1|1)$ is rational, then $\varrho[\zeta_{jk}(n)]$ are
periodic sequences. If for none of the values $n$ in a period,
$\varrho[\zeta_{jk}(n)]=0$, then beyond a certain value of $n$, $\xi_n$ is
smaller than the minimum value of $\varrho[\zeta_{jk}(n)]$ along the period and $1-\xi_n$ is larger
than the maximum, and so,
$\calI\{\varrho[\zeta_{jk}(n)]\notin(\xi_n,1-\xi_n)\}$ all vanish for
{\it all} large $n$. The expression
$$\frac{1}{M}\sum_{j=1}^r\sum_{k=1}^rp_j\pi_k
\calI\{\varrho[\zeta_{jk}(n)]\notin(\xi_n,1-\xi_n)\},$$
which generates the gap between the upper bound 
and the lower bound on $R_n$, can
be interpreted as an asymptotic approximation of 
the probability that $-\log\mu(X^n)$ falls in the vicinity
(within distance $O(\xi_n)$) of
an integer. For example, when the source is purely dyadic ($M=1$), then
$-\log\mu(X^n)$ is integer with probability 1, and indeed, the expression
in the last display is equal to 1. In this case, Theorem 1 is useless, but
it is also redundant, because in this case, we clearly know that $R_n$ vanishes. The reason
for this ``uncertainty'' around integer values of $-\log\mu(X^n)$ is that
these are the discontinuity points of the function $\varrho[-\log\mu(X^n)]$,
and in the proof of Theorem 1, 
the function $\varrho$ is expanded as a series of 
trigonometric polynomials whose convergence is
problematic in the neighborhood of discontinuities. 
Thus, we believe that the uncertainty in the
characterization of $R_n$ around these points should be attributed 
more to the
limitations of the analysis methods than to the real behavior of $R_n$.
In other words, we conjecture that, in fact, $R_n=\Omega_n+o(1)$ for {\it all}
large $n$, and not just for most large values of $n$.
It should be pointed out that
these issues were admittedly 
overlooked in \cite{merhav12} and \cite{spa00} (beyond the cases of a purely
dyadic source, which was ruled out in the first place).
The essential results therein are 
nonetheless re-confirmed here as a special case,
upon carrying out a more rigorous analysis.

The expression of the oscillatory case, $\Omega_n$, 
is not quite intuitive at first glance,
therefore, in this paragraph, we make an attempt to give some 
quick insight,
which captures the essence of the main points. The arguments here are
informal and non-rigorous (the rigorous proof is in Section 3).
The Fourier series expansion
of the periodic function $\varrho$ is given by
\begin{equation}
\varrho(u)=\frac{1}{2}+\sum_{m\ne 0}a_me^{2\pi imu}
\end{equation}
and the important fact about the coefficients is that they are
inversely proportional to $m$, so that for every two integers $k$ and $m$,
$a_{m\cdot k}=a_m/k$. Now, when computing
$R_n=\bE\{\varrho[-\log\mu(X^n)]\}$, let us take the liberty of exchanging the
order between the expectation and the summation, i.e.,
\begin{equation}
R_n=\frac{1}{2}+\sum_{m\ne 0}a_m\bE\{e^{-2\pi im\log\mu(X^n)}\}.
\end{equation}
It turns out that under the conditions of the oscillatory mode, 
$\bE\{e^{-2\pi
im\log\mu(X^n)}\}$ tends to zero as $n\to\infty$ for all $m$, except for
multiples\footnote{The convergent mode can be treated as
a special case of this statement with $M=\infty$.} of $M$, 
namely, $m=\ell M$, $l=\pm 1,\pm 2,\ldots$. Thus, for large
$n$, we have
\begin{eqnarray}
R_n&\approx&\frac{1}{2}+\sum_{\ell\ne 0}a_{\ell M}\bE\{e^{-2\pi i\ell
M\log\mu(X^n)}\}\nonumber\\
&=&\frac{1}{2}+\frac{1}{M}\sum_{\ell\ne 0}a_{\ell}\bE\{e^{-2\pi i\ell
M\log\mu(X^n)}\}\nonumber\\
&=&\frac{1}{2}+\frac{1}{M}\left\{\bE\varrho[-M\log\mu(X^n)]-\frac{1}{2}\right\}\nonumber\\
&=&\frac{1}{2}\left(1-\frac{1}{M}\right)+\frac{1}{M}\bE\varrho[-M\log\mu(X^n)].
\end{eqnarray}
Now, consider the set of all $\{x^n\}$ that begin from state $x_1=j$ and end
at state $x_n=k$. Their total probability is about $p_j\pi_k$ for large $n$
since $X_n$ is almost independent of $X_1$. 
It turns out that all these sequences
have exactly the same
value of $\varrho[-M\log\mu(x^n)]$, 
which is exactly $\varrho[\zeta_{jk}(n)]$
(or, in other words, 
$\varrho[-M\log\mu(x^n)]=\varrho[\zeta_{x_1x_n}(n)]$
independently of $x_2,\ldots,x_{n-1}$) and this
explains the expression of $\Omega_n$. 
The reason 
for this property of $\varrho[-M\log\mu(x^n)]$ is the rationality
conditions $\left<M\cdot\alpha_{uv}\right>=0$, $u,v\in\{1,2,\ldots,r\}$, which imply that
$\left<M\log p(x_t|x_{t-1})\right>=\left<M\log
[p(x_t|1)p(1|1)/p(x_{t-1}|1)]\right>$, and so, 
\begin{eqnarray}
\left<-M\log\mu(x^n)\right>&=&\left<-M\log p_j\right>+\sum_{t=2}^n\left<-M\log
p(x_t|x_{t-1})\right>~~\mbox{mod}~1\nonumber\\
&=&\left<-M\log p_j\right>+\sum_{t=2}^n\left<-M\log
[p(x_t|1)p(1|1)/p(x_{t-1}|1)]\right>~~\mbox{mod}~1
\end{eqnarray}
which, thanks to the telescopic summation, 
is easily seen to coincide with the fractional part of $\zeta_{jk}(n)$,
and of course, $\varrho[\zeta_{jk}(n)]$ depends on $\zeta_{jk}(n)$ only
via its fractional part.

Consider next the following example for using Theorem 1.

\noindent
{\bf Example 1.}
Consider a Markov source for which the
rows of
$P$ are all permutations of the first row, which is
$\bp=(p_1,\ldots,p_r)$.
Now, assuming that $\alpha_j\eqd \log(p_1/p_j)$ are all rational, let $M$ be the least
common multiple of their denominators (i.e., the common denominator)
when each one of them is expressed as a ratio between
two relatively prime integers. Then, 
\begin{eqnarray}
\varrho[\zeta_{jk}(n)]&=&\varrho[
-M(n-1)\log p(1|1)+M\log p(j|1)-M\log p(k|1)-M\log p_j]\nonumber\\
&=&\varrho[
-M(n-1)\log p_1+M\log p_j-M\log p_k-M\log p_j]\nonumber\\
&=&\varrho[
-M(n-1)\log p_1-M\log p_k]\nonumber\\
&=&\varrho(
-Mn\log p_1+M\log p_1-M\log p_k)\nonumber\\
&=&\varrho(-Mn\log p_1),
\end{eqnarray}
where in the last step, we have used the fact that $(M\log p_1-M\log p_k)$ is
integer and that $\varrho$ is a periodic function with period 1. Thus,
with the exception of the minority of `problematic' values of $n$, we have
\begin{eqnarray}
\label{e-mem}
R_{n}&=&
\frac{1}{2}\left(1-\frac{1}{M}\right)+
\frac{1}{M}\sum_{j=1}^r\sum_{k=1}^rp_j\pi_k\varrho[\zeta_{jk}(n)]+o(1)\nonumber\\
&=&\frac{1}{2}\left(1-\frac{1}{M}\right)+\frac{1}{M}
\sum_{j=1}^r\sum_{k=1}^rp_j\pi_k\varrho(-nM\log p_1)+o(1)\nonumber\\
&=&\frac{1}{2}\left(1-\frac{1}{M}\right)+\frac{1}{M}\varrho(-nM\log p_1)+o(1).
\end{eqnarray}
If not all $\alpha_j$ are rational, then $R_n\to 1/2$, as predicted
by Theorem~\ref{th-main}.
To see why the conditions of Theorem 1 lead
to the rationality condition herein, let us denote $u_{jk}=\left<m\log
[p(j|1)/p(k|1)]\right>$, and $v_{jk}=\left<m\log[p(j|j)/p(j|k)]\right>$.
Then, the conditions of Theorem 1
mean that $u_{jk}+v_{jk}=0$ and
for all pairs $j$ and $k$. Therefore,
the number of constraints here is of
the order of $r^2$, whereas the number of degrees of freedom that generate
these variables, in this example, is $r-1$, i,e., the variables
$\left<m\log(p_1/p_j)\right>$, $j=2,3,\ldots,r$.
Thus, we can think
of this as an overdetermined set
of homogeneous linear equations whose only solution is zero, meaning that
$\left<m\log(p_1/p_j)\right>$, $j=2,3,\ldots,r$, all vanish.
Note that the memoryless source is a special case of this
example, where the rows of
$P$ are all identical 
to the first row, $(p_1, \ldots, p_r)$. Indeed, eq.\ (\ref{e-mem})
coincides with the expression of the memoryless case (see 
\cite{merhav12}, \cite{spa00} and the Introduction of this paper).

\section{Proof of Theorem~\ref{th-main}}
\label{sec-proof}

\subsection{Introductory Comments}

The main idea behind the analysis of 
$R_n=\bE\{\varrho[-\log\mu(X^n)]\}$ is to approximate
the periodic function $\varrho(\cdot)$ by a sequence of trigonometric
polynomials, and then to commute the expectation with the summation and analyze
the various terms of the series. For these commutations to be legitimate, a
sufficient condition is that the
convergence would be uniform, but unfortunately, it cannot be uniform since the
function $\varrho$ is discontinuous. An alternative route that we take
is to sandwich $\varrho$ between two
continuous periodic functions, $\varrho_\theta^{-}$ and $\varrho_\theta^+$,
both with period $1$, and
both indexed by some parameter $\theta$, which when
tends to zero, the bounds become tighter and
tighter. Fej\'er's
theorem (see, e.g., \cite{Sury06}), which is the trigonometric version of the
Weierstrass theorem, provides a concrete sequence of trigonometric polynomials, which
converges uniformly to any given periodic function which is continuous.
The program of the proof is to apply Fej\'er's theorem to
$\varrho_\theta^{-}$, and $\varrho_\theta^+$, and use them to
obtain sandwich bounds on $R_n$.

\subsection{Preliminaries of the Proof}

Define the function $\varrho_\theta^-$ as
\begin{equation}
\varrho_\theta^-(u)=\left\{\begin{array}{ll}
\frac{1-\theta}{\theta}\cdot \left<u\right> & 0\le \left<u\right> < \theta\\
1- \left<u\right> & \theta\le \left<u\right> < 1\end{array}\right.
\end{equation}
and
\begin{equation}
\varrho_\theta^+(u)=
\varrho_\theta^-(u)+\Delta_\theta(u)
\end{equation}
where
\begin{equation}
\Delta_\theta(u)=\left\{\begin{array}{ll}
1-\frac{\left<u\right>}{\theta} & 0\le \left<u\right> < \theta\\
0 & \theta\le \left<u\right> < 1-\theta\\
\frac{1}{\theta}(\left<u\right>+\theta-1) & 1-\theta\le \left<u\right> <
1\end{array}\right.
\end{equation}
Obviously, $\varrho_\theta^{-}(u)$, and $\varrho_\theta^+(u)$ are continuous,
periodic functions, with period 1, and $\varrho_\theta^-(u)\le 
\varrho(u)\le\varrho_\theta^+(u)$ for every $u$.
Now, $\varrho_\theta^-$ and $\Delta_\theta$ have the following Fourier
representations:
\begin{equation}
\varrho_\theta^-(u)=\frac{1}{2}+\sum_{m\ne 0}a_m(\theta)e^{2\pi imu};~~~~
a_m(\theta)=\frac{1-e^{-2\pi im\theta}}{(2\pi im)^2\theta}
\end{equation}
and
\begin{equation}
\Delta_\theta(u)=\theta+\sum_{m\ne 0}b_m(\theta)e^{2\pi imu};~~~~
b_m(\theta)=\frac{1-\cos(2\pi m\theta)}{2\theta\pi^2m^2}.
\end{equation}
Note that for any given integers $k$ and $\ell$,
\begin{equation}
a_{\ell\cdot k}(\theta)=\frac{a_\ell(k\theta)}{k}
\end{equation}
and similarly
\begin{equation}
b_{\ell\cdot k}(\theta)=\frac{b_\ell(k\theta)}{k}.
\end{equation}
These identities will be important later on, in order to return from the
series expansions back to the original functions.
The $N$--the order F\'ejer approximations are given by
\begin{equation}
\left\{\varrho_\theta^-(u)\right\}_N\dfn\frac{1}{2}+\sum_{|m|=1}^N
a_m(\theta)\cdot\left(1-\frac{|m|}{N+1}\right)e^{2\pi imu}
\end{equation}
and
\begin{equation}
\left\{\Delta_\theta(u)\right\}_N\dfn\theta+\sum_{|m|=1}^N
b_m(\theta)\cdot\left(1-\frac{|m|}{N+1}\right)e^{2\pi imu}.
\end{equation}
According to Fej\'er's theorem, as $N\to\infty$, these functions converge
uniformly to $\varrho_\theta^-(u)$ and $\Delta_\theta(u)$, respectively. However, it
should be kept in mind that in order to guarantee that the absolute error
would be uniformly within less than a given $\epsilon$
(for all three functions $\rho_\theta^+$, $\rho_\theta^-$, and $\Delta_\theta$), 
the integer $N$ should be at
least as large as some 
$N_0(\epsilon,\theta)$ (or $N_0$ for shorthand notation), 
which grows both as $\epsilon$
decreases and as
$\theta$ decreases.
In particular, following the proof of Fej\'er's theorem \cite[p.\ 6]{Sury06}
(see also Appendix herein), it is readily seen that for all three functions,
$\rho_\theta^+$, $\rho_\theta^-$, and $\Delta_\theta$,
\begin{equation}
\label{epsilonN}
\epsilon_0(N,\theta)\dfn\inf_{0<\delta<
1/2}\left[\frac{\delta}{\theta}+\frac{1}{N\sin^2(\pi\delta)}\right]
\end{equation}
is an upper bound on the maximum approximation error when $N$ terms of the
Fej\'er series are used. Thus,
$N_0(\epsilon,\theta)$ can be defined as the smallest integer $N$
such that $\epsilon_0(N,\theta)\le\epsilon$. Obviously, by
definition
\begin{equation}
\label{obvious}
\epsilon_0[N_0(\epsilon,\theta),\theta]\le\epsilon.
\end{equation}
We will make use of this simple inequality later on.

\subsection{General Lower and Upper Bounds on $R_n$}

We proceed with some general lower and upper bounds on $R_n$.
As for the lower bound, we have
\begin{eqnarray}
R_n&=&\bE\left\{\varrho(-\log\mu(X^n))\right\}\nonumber\\
&\ge&\bE\left\{\varrho_\theta^-(-\log\mu(X^n))\right\}\nonumber\\
&\ge&\bE\left\{\frac{1}{2}+\sum_{|m|=1}^{N_0}
a_m(\theta)\cdot\left(1-\frac{|m|}{N_0+1}\right)e^{-2\pi
im\log\mu(X^n)}-\epsilon\right\}\nonumber\\
&=&\frac{1}{2}+\sum_{|m|=1}^{N_0}
a_m(\theta)\cdot\left(1-\frac{|m|}{N_0+1}\right)\bE\left\{e^{-2\pi
im\log\mu(X^n)}\right\}-\epsilon.
\end{eqnarray}
Now, clearly
\begin{equation}
\bE\left\{e^{-2\pi im\log\mu(X^n)}\right\}=
\sum_{\bx\in\calX^n}\prod_{t=1}^n
\left[p(x_t|x_{t-1})\exp\left\{-2\pi im\log p(x_t|x_{t-1})\right\}\right].
\end{equation}
Define the $r\times r$ complex matrix $A_m$ whose entries are
\begin{equation}
a_{jk}(m)=p(k|j)\exp\left[-2\pi im\log p(k|j)\right],~~~~~j,k=1,\ldots,r.
\end{equation}
Also define the $r$--dimensional column vectors
\begin{equation}
\bc_m=(p_1\exp[-2\pi im\log p_1)], \ldots, p_r\exp[-2\pi im\log p_r])^T,
\end{equation}
and $\bd=(1,1,\ldots,1)^T$, where the
superscript $T$ denotes vector/matrix transposition.
Then, it follows that
\begin{equation}
\bE\left\{e^{-2\pi im\log\mu(X^n)}\right\}=\bc_m^TA_m^{n-1}\bd.
\end{equation}
Let $\bl_{j,m}$ and $\br_{j,m}$ be, respectively,
the left eigenvector
and the right eigenvector pertaining to the eigenvalue $\lambda_{j,m}$
($j=1,2,\ldots,r$) of the matrix $A_m$. Here, we index 
the eigenvalues of $A_m$ according to a non--increasing order
of their modulus, that is,
\begin{equation}
|\lambda_{1,m}| \geq |\lambda_{2,m}| \geq \cdots \geq |\lambda_{r,m}|.
\end{equation}
Since $P$ is a stochastic matrix (so, its maximum modulus eigenvalue is 1)
and its elements are the absolute values of the corresponding elements
of $A_m$, it follows from
\cite[Theorem 8.4.5]{noble} (see also
Lemma 1 in Subsection 3.4) that $|\lambda_{1,m}|\le 1$
(and hence $|\lambda_{j,m}|\le 1$ for all $j=1,2,\ldots,r$).
Also, the sets of left-- and right eigenvectors form a bi-orthogonal system,
i.e., $\bl_{j,m}^T\br_{k,m}=0$, $j,k=1,2,\ldots,r$, $j\ne k$.
We scale these vectors such that $\bl_{j,m}^T\br_{j,m}=1$ for all
$j=1,2,\ldots,r$.
Then by the spectral representation of
matrices \cite{noble}, we have
\begin{equation}
A_m^{n-1}\bd=\sum_{j=1}^r \lambda_{j,m}^{n-1}\cdot\bl_{j,m}^T\bd\cdot
\br_{j,m},
\end{equation}
and so,
\begin{equation}
\bc_m^TA_m^{n-1}\bd
=\sum_{j=1}^r \lambda_{j,m}^{n-1}\cdot\bl_{j,m}^T\bd\cdot\bc_m^T\br_{j,m}.
\end{equation}
On substituting this back into the lower bound on $R_n$, we obtain:
\begin{equation}
\label{glb}
R_n\ge \frac{1}{2}+\sum_{|m|=1}^{N_0}
a_m(\theta)\cdot\left(1-\frac{|m|}{N_0+1}\right)\cdot
\sum_{j=1}^r \lambda_{j,m}^{n-1}\cdot\bl_{j,m}^T\bd\cdot\bc_m^T\br_{j,m}
-\epsilon.
\end{equation}
In a similar manner, we obtain the following upper bound
\begin{eqnarray}
\label{gub}
R_n&=&\bE\left\{\varrho(-\log\mu(X^n))\right\}\nonumber\\
&\le&\bE\left\{\varrho_\theta^+(-\log\mu(X^n))\right\}\nonumber\\
&=&\bE\left\{\varrho_\theta^-(-\log\mu(X^n))\right\}+
\bE\left\{\Delta_\theta(-\log\mu(X^n))\right\}\nonumber\\
&\le&\frac{1}{2}+\theta+\sum_{|m|=1}^{N_0}
[a_m(\theta)+b_m(\theta)]\cdot\left(1-\frac{|m|}{N_0+1}\right)\bE\left\{e^{-2\pi
im\log\mu(X^n)}\right\}+\epsilon\nonumber\\
&=&\frac{1}{2}+\theta+\sum_{|m|=1}^{
N_0}[a_m(\theta)+b_m(\theta)]\cdot\left(1-\frac{|m|}{N_0+1}\right)\sum_{j=1}^r
\lambda_{j,m}^{n-1}\cdot\bl_{j,m}^T\bd\cdot\bc_m^T\br_{j,m}\nonumber\\
& &+\epsilon.
\end{eqnarray}
Let us define now
\begin{eqnarray}
\gamma_n(\epsilon,\theta)&\dfn&\sum_{|m|=1}^{
N_0}[|a_m(\theta)|+|b_m(\theta)|]
\cdot\left(1-\frac{|m|}{N_0+1}\right)\sum_{j:~|\lambda_{j,m}|<1}|\lambda_{j,m}|^{n-1}|\bl_{j,m}^T{\bf
1}\cdot\bc_m^T\br_{j,m}|\nonumber\\
& &+\epsilon+\theta.
\end{eqnarray}
and recall that $N_0$ depends on $\epsilon$ and $\theta$.
Obviously, for every fixed $\epsilon$ and $\theta$,
the double sum over $m$ and $j$, in the expression of
$\gamma_n(\epsilon,\theta)$, tends to zero as $n\to\infty$
since all terms contain a factor $|\lambda_{j,m}|^{n-1}$ and by definition of
these terms, only 
$|\lambda_{j,m}|< 1$ are included in the summation.
This means that if we let $\epsilon$ and $\theta$ tend to zero slowly enough
with $n$, thus denoting them by $\epsilon_n$ and $\theta_n$, we have
$\gamma_n(\epsilon_n,\theta_n)\to 0$. In particular, let us define $\epsilon_n$
and $\theta_n$ to be the minimizers\footnote{Note that with this choice,
$\theta_n$ and $\epsilon_n$ depend only on the parameters of the source
$\mu$.} of $\gamma_n(\epsilon,\theta)$. Then, obviously,
$\gamma_n\dfn\gamma_n(\epsilon_n,\theta_n)\to 0$ as $n\to\infty$.
Then, our upper and lower bounds become
\begin{equation}
\label{glb1}
R_n\ge \frac{1}{2}+\sum_{|m|=1}^{N_0}
a_m(\theta_n)\cdot\left(1-\frac{|m|}{N_0+1}\right)\cdot
\sum_{j:~|\lambda_{j,m}|=1}\lambda_{j,m}^{n-1}\cdot\bl_{j,m}^T\bd\cdot\bc_m^T\br_{j,m}-\gamma_n,
\end{equation}
and
\begin{equation}
\label{gub1}
R_n\le \frac{1}{2}+\sum_{|m|=1}^{
N_0}[a_m(\theta_n)+b_m(\theta_n)]\cdot\left(1-\frac{|m|}{N_0+1}\right)\sum_{j:~|\lambda_{j,m}|=1}
\lambda_{j,m}^{n-1}\cdot\bl_{j,m}^T\bd\cdot\bc_m^T\br_{j,m}+\gamma_n.
\end{equation}

\subsection{Criteria for the Convergent and Oscillatory
Modes}

Considering the derived lower bound and the upper bound on $R_n$ (eqs.\
(\ref{glb1}) and (\ref{gub1}),
it is apparent that
the key issue that distinguishes between the convergent mode and the
oscillatory mode of $R_n$, is to determine under
what conditions the modulus of the dominant eigenvalue, $\lambda_{1,m}$,
namely, the {\it spectral radius} of $A_m$, denoted $\rho(A_m)$,
is equal to unity and under what conditions it is strictly less than unity
(obviously, it cannot be larger than unity). The former case is the
oscillatory mode and the latter case is the convergent one.
To this end, the following
lemma, that appears in \cite{noble} (with minor modifications in its phrasing),
and that has already been used in earlier related studies
\cite{jst01}, \cite{js12}, proves useful.

\begin{lemma} \cite[Theorem 8.4.5, p.\ 509]{noble}
\label{lem-jst}
Let $F=\{f_{kj}\}$ and $G=\{g_{kj}\}$ be two $r\times r$ matrices.
Assume that $F$ is a real,
non--negative and irreducible matrix, 
$G$ is a complex matrix, and $f_{kj}\ge |g_{kj}|$ for all
$k,j\in\{1,2,\ldots,r\}$. Then, $\rho(G)\ge \rho(F)$ 
with equality if and only if there exist
real numbers $s$, and $w_1,\ldots,w_r$ such that $G=e^{2\pi is}DFD^{-1}$,
where $D=\mbox{diag}\{e^{2\pi i w_1},\ldots,e^{2\pi i w_r}\}$.
\end{lemma}
The proof of the necessity of the condition $G=e^{2\pi is}DFD^{-1}$ appears
in \cite{noble} (see also \cite{jst01}, \cite{js12}). The sufficiency is
obvious since the matrix $DFD^{-1}$ is similar to $F$ and hence has the same
set of eigenvalues.

We wish to apply Lemma \ref{lem-jst} in order to distinguish between the two
aforementioned cases
concerning the spectral radius of $A_m$. Consider
the state transition probability matrix $P$
in the role of $F$ of Lemma \ref{lem-jst}
(i.e., $f_{kj}=p(j|k)$) and the matrix $A_m$ in the role of $G$. Since $P$
is assumed positive in this part, then it is obviously non--negative and
irreducible. Since it is a stochastic matrix,
its spectral radius is, of course, $\rho(P)=1$. Also, by definition of $A_m$,
as the matrix $\{p(j|k)\cdot\exp[-2\pi im\log p(j|k)]\}$,
it is obvious that the elements of $P$ are the absolute values of the
corresponding elements of $A_m$, and so, all the conditions of
Lemma \ref{lem-jst} clearly apply.
The lemma then tells us that $\rho(A_m)=\rho(P)=1$ if and only if there
exist real numbers $s$ and $w_1,\ldots w_r$ such that:
\begin{equation}
\label{basicondition}
-m\log p(j|k) = (s+w_k-w_j)~\mbox{mod}~1,~~~j,k=1,\ldots,r,
\end{equation}
where $x = y~\mbox{mod}~1$ means that the fractional parts of $x$ and $y$ are
equal, that is, $\left<x\right>=\left<y\right>$.

To find a vector $\bw=(w_1,\ldots,w_r)$ and a number $s$ with
this property (if exist), we take the following approach: Consider first the choice
$k=j$ in (\ref{basicondition}). 
This immediately tells us that $s$, if exists, must be equal to $-m\log
p(j|j)$ (mod $1$) for every $j=1,\ldots,r$. In other words, one set of
conditions is that $-m\log p(j|j)$ are all equal (mod $1$),
or equivalently,
\begin{equation}
\label{diagonalcond}
\left<m\log\frac{p(j|j)}{p(1|1)}\right>=0,~~~~~j=2,3,\ldots,r,
\end{equation}
and then $s$ is taken to be the common value of all $\left<-m\log
p(j|j)\right>$.
Thus, eq.\ (\ref{basicondition}) becomes
\begin{equation}
\label{condition1}
m\log\frac{p(j|j)}{p(j|k)} = (w_k-w_j)~\mbox{mod}~1,~~~j,k=1,\ldots,r,
\end{equation}
and it remains to find the vector $\bw$ if possible. To this end,
observe that if $\bw$
satisfies (\ref{condition1}), then for every constant $c$, $\bw+c$ also
satisfies (\ref{condition1}). Taking $c=-w_1$,\footnote{The choice of the
first component of $\bw$ is arbitrary.}
it is apparent that
if (\ref{condition1}) can hold for some $\bw$, then there is such a vector
whose first component vanishes, and then by setting $k=1$ in
(\ref{condition1}), we learn that
\begin{equation}
\label{wj}
w_j=\left<m\log \frac{p(j|1)}{p(j|j)}\right>, ~~~~~j=1,\ldots,r, 
\end{equation}
is a legitimate choice.
Thus, (\ref{condition1}) becomes
\begin{equation}
\label{condition2}
\left<m\log\left[\frac{p(j|1)p(j|j)}{p(k|1)p(j|k)}\right]\right>=0~~~~j,k=1,\ldots,r.
\end{equation}
Note that by setting $k=1$ in (\ref{condition2}), we get (\ref{diagonalcond})
as a special case,
which means that (\ref{condition2}), applied to all $j,k\in\{1,2,\ldots,r\}$,
are all the necessary and sufficient
conditions needed for
$\rho(A_m)=1$. Now, a necessary and sufficient condition for eq.\ (\ref{condition2})
to hold for {\it some} integer $m$, is that the numbers
\begin{equation}
\alpha_{jk}=\log \left[\frac{p(j|1)p(j|j)}{p(k|1)p(j|k)}\right]
\end{equation}
would be all rational.

We next prove the asymptotic expressions for $R_n$, first, for the case
where some $\{\alpha_{jk}\}$ are irrational, which means that $\rho(A_m) < 1$ for all
$m\ne 0$ (convergent mode), and then for the case where all $\{\alpha_{jk}\}$ are rational,
which means that there are non--zero values of $m$ for which $\rho(A_m)=1$
(oscillatory mode).

\subsection{Bounds on $R_n$ in the Convergent and Oscillatory Modes}

When some $\alpha_{jk}$ are irrational, then for all
$m\ne 0$ and $j\in\{1,2,\ldots,r\}$, we have $|\lambda_{j,m}|<1$, and so, the second terms
(i.e., the sums over $m$) in eqs.\ (\ref{glb1}) and (\ref{gub1})
do not exist. Consequently, we immediately get $R_n\ge \frac{1}{2}-\gamma_n$ and
$R_n\le \frac{1}{2}+\gamma_n$, namely, $R_n=\frac{1}{2}+o(1)$.

Consider now the case where all $\{\alpha_{jk}\}$ are rational, and so,
there exist $m\ne 0$ with $\rho(A_m)=1$.
Our first step is to establish the fact that if $M$
is the smallest positive integer
$m$ that satisfies (\ref{condition2}), then any other non--zero integer $m$ satisfies
this property if and only if
it is an integral multiple of $M$. The fact that integer
multiples of $M$ satisfy (\ref{condition2}) is obvious since $\left<k\cdot
M\alpha_{jk}\right>=\left<k\cdot\left<M\alpha_{jk}\right>\right>=\left<k\cdot
0\right> =0$. To see why the converse is true as well, let $M'$ be another
integer satisfying (\ref{condition2}). If $M'$ is not an integer multiple of
$M$, it must be larger than $M$ since $M$ was defined as the smallest integer
satisfying (\ref{condition2}). Now, if $M$ and $M'$ both satisfy
(\ref{condition2}), then so does $M''=M'-\lfloor M'/M\rfloor\cdot M$, but
$M''$ must be strictly smaller than $M$, which is a contradiction.

This means that for $m=\ell M$, $\ell=\pm 1,\pm 2,\ldots$,
and only for these integers, $A_m$ has a modulus 1 eigenvalue
\begin{equation}
\lambda_{1,\ell M}=\exp\left[2\pi i\left<-\ell M\log p(1|1)\right>\right]=
\exp\left[-2\pi i\ell M\log p(1|1)\right]
\end{equation}
and the corresponding vector $\bw$ is $\ell$ times (mod $1$) 
the vector $\bw$ associated with $m=M$.
By the Perron--Frobenius theorem \cite{noble},
all other eigenvalues 
have modulus strictly less than 1, and they will contribute 
exponentially small terms to $R_n$. 
Since $\lambda_{1,\ell M}^{-1}A_{\ell M}$ is similar to $P$, under the transformation
matrix $D=\mbox{diag}\{e^{2\pi i w_1},\ldots,e^{2\pi i w_r}\}$, $w_j=\left<\ell
M\log[p(j|1)/p(j|j)]\right>$, $j=1,2,\ldots,r$
(see Lemma \ref{lem-jst}), then by (\ref{wj}),
the right- and left eigenvectors associated with $\lambda_{1,\ell M}$ are,
respectively,
\begin{equation}
\br_{1,\ell M}=D\cdot\bd=\left(1, e^{2\pi i\ell M\log[p(2|1)/p(2|2)]}
,\ldots,e^{2\pi i\ell M\log[p(r|1)/p(r|r)]}\right)^T, 
\end{equation}
and
\begin{equation}
\bl_{1,\ell M}=(\pi_1,\ldots,\pi_k)\cdot D^{-1}
=\left(\pi_1, \pi_2e^{-2\pi i\ell M\log
[p(2|1)/p(2|2)]},\ldots,
\pi_re^{-2\pi i\ell M \log[p(r|1)/p(r|r)]}\right). 
\end{equation}
Thus, the dominant term in $\bc_{\ell M}^TA_{\ell M}^{n-1}\bd$ becomes:
\begin{equation}
\lambda_{1,\ell M}^{n-1}\cdot\bl_{1,\ell M}^T\bd\cdot\bc_{\ell M}^T\br_{1,\ell
M}=\sum_{j,k}p_j\pi_ke^{2\pi
i\ell\zeta_{jk}(n)},
\end{equation}
where $\zeta_{jk}(n)$ is defined as in Theorem 1.
Combining this relation with eq.\ (\ref{glb1}), $R_n$ is further lower bounded as follows:
\begin{eqnarray}
\label{longchain}
R_n&\ge&\frac{1}{2}+\sum_{|\ell|=1}^{
\lfloor N_0/M\rfloor}a_{\ell M}(\theta_n)\cdot\left(1-\frac{|\ell
M|}{N_0+1}\right)\cdot
\sum_{j,k}p_j\pi_k e^{2\pi i\ell\zeta_{jk}(n)}
-\gamma_n\nonumber\\
&=&\frac{1}{2}+\frac{1}{M}\sum_{|\ell|=1}^{
\lfloor N_0/M\rfloor}a_{\ell}(M\theta_n)\cdot\left(1-\frac{|\ell
M|}{N_0+1}\right)\cdot
\sum_{j,k}p_j\pi_k e^{2\pi i\ell\zeta_{jk}(n)}-\gamma_n\nonumber\\
&=&\frac{1}{2}+\frac{1}{M}\sum_{|\ell|=1}^{
\lfloor N_0/M\rfloor}a_{\ell}(M\theta_n)\cdot\left(1-\frac{|\ell|}{\lfloor
N_0/M\rfloor +1}\right)\cdot
\sum_{j,k}pj\pi_k e^{2\pi i\ell\zeta_{jk}(n)}-\nonumber\\
& &\frac{1}{M}\sum_{|\ell|=1}^{
\lfloor
N_0/M\rfloor}a_{\ell}(M\theta_n)\cdot\left[\frac{|\ell|}{(N_0+1)/M}-\frac{|\ell|}{\lfloor
N_0/M\rfloor +1}\right]\cdot
\sum_{j,k}p_j\pi_k e^{2\pi i\ell\zeta_{jk}(n)}
-\gamma_n\nonumber\\
&\ge&
\frac{1}{2}+\frac{1}{M}\sum_{j,k}p_j\pi_k\left\{\varrho_{M\theta_n}^-[\zeta_{jk}(n)]-\frac{1}{2}-
\eta_n\right\}-\nonumber\\
& &\frac{1}{M}\sum_{|\ell|=1}^{
\lfloor
N_0/M\rfloor}a_{\ell}(M\theta_n)\cdot\left[\frac{|\ell|}{(N_0+1)/M}-\frac{|\ell|}{\lfloor
N_0/M\rfloor +1}\right]\cdot
\sum_{j,k}p_j\pi_k e^{2\pi i\ell\zeta_{jk}(n)}-\gamma_n\nonumber\\
&\ge&\frac{1}{2}\left(1-\frac{1}{M}\right)+\frac{1}{M}
\sum_{j,k}p_j\pi_k\varrho[\zeta_{jk}(n)]
-\frac{1}{M}\sum_{j,k}p_j\pi_k\Delta_{M\theta_n}[\zeta_{jk}(n)]-\nonumber\\
& &\frac{1}{M}\sum_{j,k}p_j\pi_k\sum_{|\ell|=1}^{
\lfloor
N_0/M\rfloor}a_{\ell}(M\theta_n)\cdot\left[\frac{|\ell|}{(N_0+1)/M}-\frac{|\ell|}{\lfloor
N_0/M\rfloor +1}\right]\cdot
e^{2\pi i\ell\zeta_{jk}(n)}-\nonumber\\
& &\gamma_n-\frac{\eta_n}{M},
\end{eqnarray}
where $\eta_n$ is defined as the maximum approximation error of the function
$\varrho_{M\theta_n}^-$ using $\lfloor N_0(\epsilon_n,\theta_n)/M\rfloor$
terms of the Fej\'er series. We wish to show now that $\eta_n\to 0$ as $n\to\infty$.
Let us assume that $\epsilon_n$ and $\theta_n$ are small enough to make
$N_0=N_0(\epsilon_n,\theta_n)$ not smaller than $2M$, and so,
$\lfloor N_0/M\rfloor \ge N_0/M-1\ge N_0/2M$.
Then, using eq.\ (\ref{epsilonN}),
\begin{eqnarray}
\eta_n&\le&\epsilon_0\left[\frac{N_0(\epsilon_n,\theta_n)}{2M},M\theta_n\right]\nonumber\\
&=&\inf_{0<\delta<
1/2}\left[\frac{\delta}{M\theta_n}+\frac{2M}{N_0(\epsilon_n,\theta_n)\sin^2(\pi\delta)}\right]\nonumber\\
&<&\inf_{0<\delta<
1/2}\left[\frac{2M\delta}{\theta_n}+\frac{2M}{N_0(\epsilon_n,\theta_n)\sin^2(\pi\delta)}\right]\nonumber\\
&=&2M\cdot\epsilon_0[N_0(\epsilon_n,\theta_n),\theta_n]\nonumber\\
&\le&2M\epsilon_n\to 0,
\end{eqnarray}
where the last inequality follows from eq.\ (\ref{obvious}). Thus, $\eta_n/M$
in the last line of (\ref{longchain}), is upper bounded by $2\epsilon_n$.
The first two terms in the last expression of (\ref{longchain}) form
$\Omega_n$, as defined in Theorem 1.
Now, for the absolute value of the fourth term, it is first observed that upon
a standard algebraic manipulation under the assumption $N_0\ge 2M$, we have
\begin{eqnarray}
\bigg|\frac{1}{(N_0+1)/M}-\frac{1}{\lfloor N_0/M\rfloor+1}\bigg|&=&
\frac{M|\left<N_0/M\right>+1/M-1|}{(N_0+1)(\lfloor N_0/M\rfloor+1)}\nonumber\\
&\le&\frac{2M^2}{N_0^2}.
\end{eqnarray}
Thus, the fourth term of (\ref{longchain}) is upper bounded by the weighted
sum (with weights $p_j\pi_k$ for each pair $(j,k)$) of terms,
that are bounded as follows:
\begin{eqnarray}
\label{smallaterm}
& &\bigg|\frac{1}{M}\sum_{|\ell|=1}^{
N_0/M}a_{\ell}(M\theta_n)\cdot\left[\frac{|\ell|}{(N_0+1)/M}-\frac{|\ell|}{
N_0/M +1}\right]\cdot
e^{2\pi i\ell\zeta_{jk}(n)}\bigg|\nonumber\\
&\le&\frac{4M^2}{N_0^2}\sum_{\ell=1}^
{\lfloor N_0/M\rfloor}\ell\cdot |a_{\ell}(M\theta_n)|\nonumber\\
&=&\frac{2M}{\pi N_0^2}\sum_{\ell=1}^
{\lfloor N_0/M\rfloor }\frac{|1-e^{-2\pi i\ell M\theta_n}|}{2\pi\ell M\theta_n}\nonumber\\
&=&\frac{2M}{\pi N_0^2}\sum_{\ell=1}^
{\lfloor N_0/M\rfloor}\sqrt{\frac{2[1-\cos(2\pi\ell M\theta_n)]}{(2\pi\ell M\theta_n)^2}}.
\end{eqnarray}
To bound the summand of the last expression, consider the following:
For every positive $t$, clearly, $\sin t\le t$, and so, for every $\alpha>0$,
\begin{equation}
1-\cos\alpha=\int_0^\alpha\sin t\mbox{d}t\le\int_0^\alpha
t\mbox{d}t=\frac{\alpha^2}{2},
\end{equation}
which for $\alpha=2\pi\ell M\theta_n$, 
implies that the summand is bounded by $1$, and hence
the expression in the last chain of inequalities is further upper bounded by
$\delta_n\dfn 2/(\pi N_0)$.
Since $N_0=N_0(\epsilon_n,\theta_n)\to \infty$, then $\delta_n\to 0$,
and we have
\begin{eqnarray}
R_n&\ge&
\frac{1}{2}\left(1-\frac{1}{M}\right)+\frac{1}{M}
\sum_{j,k}p_j\pi_k\varrho[\zeta_{jk}(n)]
-\frac{1}{M}\sum_{j,k}p_j\pi_k\Delta_{M\theta_n}[\zeta_{jk}(n)]-\gamma_n-2\epsilon_n-\delta_n\nonumber\\
&\ge&\frac{1}{2}\left(1-\frac{1}{M}\right)+\frac{1}{M}
\sum_{j,k}p_j\pi_k\varrho[\zeta_{jk}(n)]-\nonumber\\
& &\frac{1}{M}\sum_{j,k}p_j\pi_k\calI\{\varrho[\zeta_{jk}(n)]\notin(M\theta_n,1-M\theta_n)\}-
\gamma_n-2\epsilon_n-\delta_n,
\end{eqnarray}
and so, the lower bound of Theorem 1 is obtained with $\xi_n\dfn
M\theta_n$. In the very same manner, the upper bound on $R_n$ 
is given by
\begin{eqnarray}
R_n&\le&
\frac{1}{2}\left(1-\frac{1}{M}\right)+
\frac{1}{M}\sum_{j,k}p_j\pi_k\varrho_{M\theta_n}^-[\zeta_{jk}(n)]+\nonumber\\
& &\frac{1}{M}\sum_{j,k}p_j\pi_k\Delta_{M\theta_n}[\zeta_{jk}(n)]+\gamma_n+2\epsilon_n+\delta_n\\
&\le&\frac{1}{2}\left(1-\frac{1}{M}\right)+
\frac{1}{M}\sum_{j,k}p_j\pi_k\varrho[\zeta_{jk}(n)]+\nonumber\\
& &\frac{1}{M}\sum_{j,k}p_j\pi_k\calI\{\varrho[\zeta_{jk}(n)]
\notin(\xi_n,1-\xi_n)\}+\gamma_n+2\epsilon_n+\delta_n,
\end{eqnarray}
which is the upper bound of Theorem 1. Here,
one has to bound also an expression similar to (\ref{smallaterm}), but with
$a_{\ell}(M\theta_n)$ being replaced by 
$b_{\ell}(M\theta_n)$, and the bounding technique is similar.
This completes the proof of Theorem 1.

\section{Extensions}
\label{sec-extension}

We now discuss some extensions of Theorem~\ref{th-main}. In particular,
we drop the assumption that all transition probabilities must be strictly
positive and first assume that $P$ corresponds to an irreducible aperiodic Markov source.
Then we drop the aperiodicity constraint.

\subsection{Irreducible Aperiodic Markov Sources}

When some of the entries of the matrix $P$ vanish, then obviously,
Theorem 1 cannot be used as is since the corresponding parameters
$\alpha_{jk}$ are
no longer well defined. 
Lemma 1, which stands at the heart of the proof of
Theorem 1, can still be used as long as $P$ is irreducible, 
but more caution should be exercised.
The key issue is still to determine
whether there exist parameters $s$ and $\bw$ (and to find them if exist) that satisfy 
\begin{equation}
\label{eqs}
-m\log p(j|k) = (s +w_k-w_j)~\mbox{mod}~1,
\end{equation}
but now these equations are imposed only for the pairs $(j,k)$ for which
$p(j|k)>0$ (as for the other pairs $a_{jk}(m)=p(j|k)=0$ satisfy the conditions
of Lemma 1 automatically anyway). The approach taken in the 
solution for $s$ and $\bw$, that was derived
in the first part of Section 3, can still be applied, with some minor
modifications, as long as at least some particular subsets of the entries of $P$ are
still positive. 

For example, if one or more diagonal element of $P$ is
positive, and for all positive $p(j|j)$, the numbers $\left<-m\log p(j|j)\right>$ are
equal, then $s$ can still be taken to be the common value of all these
numbers. If, in addition, at least one row of $P$ is
strictly positive, say, row number $l$, then $w_j$ can be taken to be
$\left<m\log[p(l|l)/p(j|l)]\right>$, and then the rationality condition
of Theorem 1 is replaced by the condition that
\begin{equation}
\alpha_{jk}'=\log\left[\frac{p(j|l)p(l|l)}{p(k|l)p(j|k)}\right]
\end{equation}
must be rational for all $(j,k)$ with $p(j|k)>0$. The bounds on $R_n$ in the
oscillatory mode
would be exactly as in Theorem 1, but with the above assignments of $s$ and
$\bw$.

For a general non-negative matrix $P$, however, 
it may not be a trivial task to determine whether equations
(\ref{eqs}) have a solution, and if so, what this solution is. In fact, it may
be simpler and more explicit to check directly if $A_m$ 
has an eigenvalue on the unit circle (which
thereby dictates $s$) and then to find $\bw$ using Lemma 1. This would lead
to the following generalized version of Theorem 1.

\begin{theorem}
\label{th-ext}
Consider the Shannon code of block length $n$ for an irreducible aperiodic 
Markov source. Let $M$ be defined as the smallest positive integer $m$
such that
\begin{equation}
\label{condthm2}
\rho(A_m)\equiv |\lambda_{1,m}|=1
\end{equation}
and as $M=\infty$ if (\ref{condthm2}) 
does not hold for any positive integer $m$.
Then, $R_n$ is characterized as follows:\\
(a) If $M=\infty$, then 
\begin{equation}
R_n=\frac{1}{2}+o(1).
\end{equation}
(b) If $M<\infty$, then the bounds of Theorem 1, part (b), hold with
$\zeta_{jk}(n)$ being redefined according to
\begin{equation}
\label{e5}
\zeta_{jk}(n)=
M[(n-1)s+w_j-w_k-\log p_j],
\end{equation}
where
\begin{equation}
s=\frac{\mbox{\rm arg}\{\lambda_{1,M}\}}{2\pi}
\end{equation}
and 
\begin{equation}
w_j =\frac{\mbox{\rm arg}\{x_j\}}{2\pi},~~~~j=1,2,\ldots,r,
\end{equation}
$x_j$ being the $j$--th component of the right eigenvector $\bx$ of $A_M$,
which is associated with the dominant eigenvalue $\lambda_{1,M}$.
\end{theorem}
The proof of Theorem 2 is very similar to that of Theorem
1, and hence we will not provide it here. In a nutshell, 
we observe that the Perron--Frobenius Theorem and Lemma \ref{lem-jst}
are still applicable. Then, we use
the necessity of the condition $A_m=e^{2\pi is}DPD^{-1}$ and the fact that
once this condition holds, the vector $\bx=D\cdot\bd=(e^{2\pi iw_1},\ldots,
e^{2\pi iw_r})^T$ is the right eigenvector
associated with the dominant eigenvalue $\lambda_{1,m}=e^{2\pi is}$.

Unfortunately, Theorem 2 does not suggest a practical
way to find $M$.
One must start with $m=1$, check if $\rho(A_1)=1$; if not -- increment $m$ to
$2$, check $\rho(A_2)$, and so on. In the event that $M=\infty$, we do not have a
stopping rule and we may keep incrementing $m$ indefinitely.
An interesting point to note, however, is that the oscillatory
expression goes to $1/2$ when $M$ grows without bound. This means that given
the block length $n$, it is sufficient to stop incrementing $m$ at some $m(n)$,
where $m(n)$ is an arbitrary function that
grows (and no matter how slowly) with $n$. This is because the oscillatory
expression will then be $1/2+o(n)$
anyway, just like the convergent expression, so the
distinction between the two modes looses its meaning.

Finally, it is instructive to demonstrate an example of a {\it reducible} Markov
source, for which Theorems 1 and 2 do not hold, and see that even in a simple
situation ($r=2$), once the
irreducibility assumption is dropped, the two--mode behavior, predicted by
Theorems 1 and 2, disappears.
Thus, the point in Example 2 below is that the irreducibility assumption is not imposed just for technical
convenience. It is actually essential for Theorems 1 and 2 to hold.
\med
{\bf Example 2.} {\it Reducible Markov source}.
Consider the case $r=2$, where $p(1|2)=0$
and $\alpha\eqd p(2|1)\in(0,1)$,
i.e.,
\begin{equation}
P=\left(\begin{array}{cc}
1-\alpha & \alpha\\
0 & 1 \end{array}\right).
\end{equation}
Assume also that $p_1=1$ and $p_2=0$.
Since this is a reducible Markov source (once in state 2, there is no way back
to state 1), we cannot use Theorems 1 and 2, but
we can still find an asymptotic expression of the redundancy in a direct manner:
Note that the chain starts at state `1' and
remains there for a random duration, which is a geometrically distributed
random variable with parameter $(1-\alpha)$. Thus, the probability of $k$ 1's
(followed by $n-k$ 2's) is about
$(1-\alpha)^k\cdot\alpha$ (for large $n$) and so the argument of the function
$\varrho(\cdot)$
should be the negative
logarithm of this probability. Taking the expectation w.r.t.\ the randomness
of $k$, we readily have
\begin{equation}
R_n=\sum_{k=0}^\infty\alpha(1-\alpha)^k
\varrho[-\log\alpha
-k\log(1-\alpha)]+o(1).
\end{equation}
We see then that there is
{\it no oscillatory mode} in this case, as
$R_n$ always tends to a constant
that depends on $\alpha$, in contrast to
the convergent mode of Theorems 1 and 2, where
the limit is always $1/2$, independently
of the source statistics. To summarize, it is observed that the behavior here is very
different from that of the irreducible case, characterized by Theorems 1 and
2.

\subsection{Irreducible Periodic Markov Sources}

Consider now an irreducible periodic Markov source. The
Perron-Frobenius theorem and Lemma \ref{lem-jst} still hold \cite{noble}. 
However, the matrix
$P$ now has $d$ eigenvalues on the unit circle, namely, all the $d$--th
roots of unity \cite{noble}, where $d$ is the period, i.e., 
\begin{equation}
\lambda_t'=e^{2\pi i t/d}, ~ ~ ~ ~ t=0,1,\ldots,d-1. 
\end{equation}
Let $\br_t$ and $\bl_t$ be the right- and the left
eigenvectors of $P$ that are associated with $\lambda_t'$.
The analysis is similar as in the aperiodic case,
except that we now have $d$ oscillatory terms, one for each eigenvalue on the
unit circle. 
Indeed, suppose that for
some $m$, the matrix $A_m$ has a modulus--1 eigenvalue $\lambda=e^{2\pi is}$.
Then, of course, 
\begin{equation}
A_m'\eqd e^{-2\pi is}A_m
\end{equation}
has eigenvalue 1. By definition, the entries of $P$ are still the absolute
values of the corresponding entries of $A_m'$,
as in Lemma~\ref{lem-jst}. Thus, by this lemma, $A_m'$ is similar to $P$, and
so it has the same eigenvalues as $P$. Among them,
the $d$--th roots of unity $\lambda_t'$, $t=0,1,\ldots,d-1$ are eigenvalues of
$A_m'$. Therefore, $A_m$
has the following eigenvalues on the unit circle: 
\begin{equation}
\lambda_{t,m}=e^{2\pi i(s+t/d)}, ~ ~ ~  t=0,1,\ldots,d-1. 
\end{equation}
Let us relabel, if necessary, the eigenvalues
of $A_m$ such that $s\in[0,1/d)$. This means that the definition of $s$ in Theorem 2 should be 
restricted to the half open interval $[0,1/d)$. Thus,
Theorem~\ref{th-ext} holds except that $\zeta_{jk}(n)$ are replaced by
\begin{equation}
\zeta_{jkt}(n)\dfn M\left[(n-1)\left(s+\frac{t}{d}\right)+w_j-w_k-\log
p_j\right],~~~~j,k\in\{1,2,\ldots,r\},~t=0,1,\ldots,d-1
\end{equation}
and the double summations over $(j,k)$ with weights $p_j\pi_k$, are
replaced by corresponding triple summations over $(j,k,t)$ with weights
$p_jr_{t,j}l_{t,k}$,
where $l_{t,k}$ is the $k$--th component of $\bl_t$ and
$r_{t,j}$ is the $j$--th component of $\br_t$. Note that $r_{0,j}=1$ and
$l_{0,k}=\pi_k$, so for $d=1$ we
indeed obtain the expression (\ref{e5})  of the aperiodic case
as a special case.

\section*{Appendix}
\renewcommand{\theequation}{A.\arabic{equation}}
    \setcounter{equation}{0}

In this appendix, we establish the relation (\ref{epsilonN}). As is shown in
\cite{Sury06},
the coefficients of the $N$--th order 
Fej\'er series expansion, $\{f(u)\}_N$, of a general periodic function $f(u)$, with period
1, are given by the Fourier coefficients $f_m$ multiplied by the ``triangular
window'' $1-|m|/(N+1)$. This means that in the original $u$-domain, the
reconstruction $\{f(u)\}_N$ is given by the convolution between $f(u)$ and the kernel
\begin{equation}
K_N(u)=\sum_{m=-N}^{N}\left(1-\frac{|m|}{N+1}\right)e^{2\pi
imu}=\frac{\sin^2[(N+1)\pi u]}{(N+1)\sin^2(\pi u)}.
\end{equation}
Since $\int_{-1/2}^{+1/2}K_N(u)\mbox{d}u=1$, we have
\begin{eqnarray}
|f(u)-\{f(u)\}_N|&=&\bigg|f(u)-\int_{-1/2}^{+1/2}\mbox{d}tf(u-t)K_N(t)\bigg|\nonumber\\
&=&\bigg|\int_{-1/2}^{+1/2}\mbox{d}t[f(u)-f(u-t)]K_N(t)\bigg|\nonumber\\
&\le&\int_{-1/2}^{+1/2}\mbox{d}t|f(u)-f(u-t)|\cdot K_N(t)\nonumber\\
&=&\int_{|t|\le\delta}\mbox{d}t|f(u)-f(u-t)|\cdot K_N(t)+\nonumber\\
& &\int_{\delta\le |t|\le 1/2}\mbox{d}t|f(u)-f(u-t)|\cdot K_N(t)
\end{eqnarray}
for every $\delta\in (0,1/2)$.
Now, in our case, for all three functions, $|t|\le\delta$ implies
$|f(u)-f(u-t)|\le\delta/\theta$, since the maximum absolute slope of all three
of them is $1/\theta$. Since $K_N(t)\ge 0$ and 
$\int_{-1/2}^{1/2}\mbox{d}tK_N(t)=1$,
the first integral in the last line is bounded by $\delta/\theta$.
As for the second integral, in our case, $|f(u)-f(u-t)|\le 1$ for all three
functions. Since the sine function is monotonically
increasing in the range $[0, \pi/2]$, then $1/2\ge |t|\ge\delta$ implies
\begin{equation}
K_N(t)\le \frac{1}{(N+1)\sin^2(\pi\delta)}< \frac{1}{N\sin^2(\pi\delta)}.
\end{equation}
Thus, for every $\delta\in(0,1/2)$,
\begin{equation}
|f(u)-\{f(u)\}_N|\le \frac{\delta}{\theta}+\frac{1}{N\sin^2(\pi\delta)}
\end{equation}
and eq.\ (\ref{epsilonN}) is obtained upon minimizing the r.h.s.\ over
the free parameter $\delta$.


\begin{thebibliography}{99}
\bibitem{abrahams}
J.~Abrahams,
``Code and parse trees for lossless source encoding,''
{\it Communications in Information and Systems}, vol.\ 1, no.\ 2,
pp.\ 113--146, April 2001.

\bibitem{blackwell}
D.~Blackwell and J.~L.~Hodges, ``The probability in the extreme tail
of a convolution,'' {\em Ann.\ Math.\ Stat.}, vol.\ 30, pp.\ 1113--1120, 1959.




\bibitem{ds04}
M.~Drmota and W.~Szpankowski,
``Precise minimax redundancy and regret,''
{\it IEEE Trans.\ Inform.\ Theory}, vol.\ 50, no.\ 11, 
pp.\ 2686--2707, November 2004.

\bibitem{esseen}
C.-G.~Esseen, ``Fourier analysis of distribution functions,''
{\em Acta Mathematica}, vol.\ 77, pp.\ 1-125, 1945.

\bibitem{fs-book}
P.~Flajolet and R.~Sedgewick, {\it Analytic Combinatorics},
Cambridge University Press, Cambridge, 2009.

\bibitem{Gallager78}
R.~G.~Gallager, ``Variations on the theme by Huffman,''
{\em IEEE Trans.~Inform.~Theory\/},
vol.~IT--24, no.~6, pp.~668--674, November 1978.

\bibitem{gray}
R.~M.~Gray,
``Quantization noise spectra,''
{\it IEEE Trans.\ Inform.\ Theory}, vol.\ 36, 
no.\ 6, pp.\ 1220-1244, November 1990.

\bibitem{noble}
R.~A.~Horn and C.~R.~Johnson,
{\it Matrix Analysis}, Cambridge University Press, Cambridge, 1985.


\bibitem{jst01}
P.~Jacquet, W.~Szpankowski, and J.~Tang,
``Average profile of the Lempel--Ziv parsing scheme for a Markovian source,''
{\it Algorithmica}, vol.\ 31, pp.\ 318--360, 2001.

\bibitem{js04}
P.~Jacquet and W.~Szpankowski,
``Markov types and minimax redundancy for Markov sources,
{\it IEEE Trans.\ Inform.\ Theory}, vol.\ 50, no.\ 7, 
pp.\ 1393--1402, July 2004.

\bibitem{js12}
P.~Jacquet and W.~Szpankowski, 
``Joint string complexity for Markov sources,''
{\it 23rd International Meeting on Probabilistic, Combinatorial and
Asymptotic Methods for the Analysis of Algorithms}, AofA'12,
{\it DMTCS Proc.}, pp.\ 303--322, Montreal, 2012.

\bibitem{KN74}
L.~Kuipers and H.~Niederreiter, {\it Uniform Distribution of Sequences},
Wiley, New York, 1974.


\bibitem{LS97}
G.~Louchard and W.~Szpankowski, ``Average redundancy of the
Lempel--Ziv code,'' {\em IEEE Trans.~Inform.~Theory\/},
vol.~43, no.~1, pp.~2--8, January 1997.

\bibitem{merhav12}
N.~Merhav,
``Relations between redundancy patterns of the Shannon code and
wave diffraction patterns of partially disordered media,''
{\it IEEE Trans.\ Inform.\ Theory}, vol.\ 58, no.\ 6, pp.\ 3402--3406, 
June 2012.

\bibitem{rissanen}
J.~Rissanen,
``Complexity of strings in the class of Markov sources,''
{\it IEEE Trans.\ Inform.\ Theory}, vol.\ IT--32, no.\ 4,
pp.\ 526--532, July 1986.

\bibitem{Savari98}
S.~A.~Savari, ``Variable--to--fixed length codes for predictable sources,''
{\it Proc.\ Data Compression Conference (DCC)}, Snowbird, UT, pp.\ 481--490,
1998.

\bibitem{SG97}
S.~A.~Savari and R.~G.~Gallager, ``Generalized Tunstall codes for sources with
memory,'' {\em IEEE Trans.~Inform.~Theory\/},
vol.~43, no.~2, pp.~658--668, March 1997.

\bibitem{spa00}
W.~Szpankowski, ``Asymptotic average redundancy of Huffman (and other) 
block Codes,''
{\it IEEE Trans.\ Inform.\ Theory}, vol.\ 46, no.\ 6, 
pp.\ 2434--2443, November 2000.

\bibitem{spa-book}
W.~Szpankowski,
{\it Average Case Analysis of Algorithms on Sequences},
John Wiley \& Sons, New York, 2001.

\bibitem{Sury06}
B.~Sury, ``Weierstrass's Theorem -- Leaving No Stone Unturned,''
{\it Workshop on Linear Algebra and Analysis}, University of Hyerabad, 2006.
http://www.isibang.ac.in/$\sim$sury/hyderstone.pdf

\end{thebibliography}
\end{document}